 \documentclass[final, 3p, times]{elsarticle}

\usepackage{amssymb}
\usepackage{lipsum}
\usepackage[utf8]{inputenc} 
\usepackage[T1]{fontenc}    
\usepackage{url}            
\usepackage{booktabs}       
\usepackage{amsfonts}       
\usepackage{nicefrac}       
\usepackage{microtype}      
\usepackage{xcolor}         
\usepackage{amsmath}
\usepackage{bm}
\newcommand\norm[1]{\left\lVert#1\right\rVert}

\usepackage{amssymb}
\usepackage{mathtools}
\usepackage{amsthm}
\usepackage{multirow}
\usepackage{graphicx}
\usepackage[hidelinks]{hyperref} 
\usepackage{calc}
\usepackage{cancel}
\journal{Computer Methods in Applied Mechanics and Engineering}

\begin{document}

\begin{frontmatter}



\title{Godunov Loss Functions for Modelling of Hyperbolic Conservation Laws}


\author[first]{Rami Cassia}
\author[first]{Rich Kerswell}
\affiliation[first]{organization={Department of Applied Mathematics and Theoretical Physics, University of Cambridge},
            addressline={Wilberforce Rd}, 
            city={Cambridge},
            postcode={CB3 0WA}, 
            country={UK}}

\begin{abstract}
Machine learning techniques are being used as an alternative to traditional numerical discretization methods for solving hyperbolic partial differential equations (PDEs) relevant to fluid flow. Whilst numerical methods are higher fidelity, they are computationally expensive. Machine learning methods on the other hand are lower fidelity but can provide significant speed-ups. The emergence of physics-informed neural networks (PINNs) in fluid dynamics has allowed scientists to directly use PDEs for evaluating loss functions. The downfall of this approach is that the differential form of systems is invalid at regions of shock inherent in hyperbolic PDEs such as the compressible Euler equations. To circumvent this problem we propose the Godunov loss function: a loss based on the finite volume method (FVM) that crucially incorporates the flux of Godunov-type methods. These Godunov-type methods are also known as approximate Riemann solvers and evaluate intercell fluxes in an entropy-satisfying and non-oscillatory manner, yielding more physically accurate shocks. Our approach leads to superior performance compared to standard PINNs that use regularized PDE-based losses as well as FVM-based losses, as tested on the 2D Riemann problem in the context of time-stepping and super-resolution reconstruction.
\end{abstract}




\end{frontmatter}

\section*{Highlights}
\begin{itemize}
    \item Introduction of Godunov loss functions which inherently incorporate solution of Riemann problems for ML modelling of compressible flow discontinuities.
    \item Godunov loss yields superior performance to baseline losses in forward and inverse 2D Riemann problems, in terms of accuracy and stability.
    \item Novel implementation of positivity preservation as a hard constraint in ML algorithms.
\end{itemize}




\section{Introduction}

Machine learning (ML) has become an increasingly popular tool in computational fluid dynamics (CFD). Applications broadly fall into the areas of CFD acceleration \cite{bar2019learning,li2020fourier, kochkov2021machine}, reduced-order modelling \cite{noack2003hierarchy,kaiser2014cluster,murata2020nonlinear}, turbulence modelling \cite{ling2016reynolds, wu2018physics, beck2019deep}, active flow control \cite{rabault2020deep, bieker2020deep, bhattacharjee2020data} and flow reconstruction \cite{fukami2019super, fukami2022machine, kelshaw2022physics}. In this work we are interested in using ML to model compressible flows governed by hyperbolic partial differential equations (PDEs) which are of relevance in aerodynamics and astrophysics. ML models in this context must ensure predictions are entropy-satisfying and prevent artificial oscillations forming in the flow. 

\citet{raissi2019physics} propose a method of integrating physics into ML frameworks by including physics-based PDE losses rather than using standard ML losses. They demonstrate the feasibility of these physics-informed neural networks (PINNs) on Navier-Stokes, Korteweg-De Vries, and Burgers' equations. The use of PINNs for modelling of physical phenomena has gained popularity in recent times, particularly in fluid dynamics. This is in part due to the added interpretability that physical losses bring to ML. Numerous architectural modifications to standard fully-connected PINNs (FC-PINNs) have been proposed, such as Fourier neural operator learning \cite{li2020fourier} and MeshGraphNets \cite{pfaff2020learning}. Additionally, \citet{ren2022phycrnet} use a convolutional long short-term memory network (Conv-LSTM) with hard imposition of initial/boundary conditions (I/BCs) to model the viscous 2D Burgers' equations. Their work demonstrates improvement over FC-PINNs due to the ability of convolutional LSTMs in capturing spatiotemporal flow features. The hard imposition of I/BCs via padding negates the use of penalty terms which reduces hyperparameter tuning. Autoregressive and residual connections of the architecture also simulate time marching. The limitation of using PDE-based PINN losses for hyperbolic systems is their invalidity in potential shock regions. \citet{mao2020physics} alleviate this limitation by clustering training points in regions where shocks develop. This however requires prior knowledge of shock location. \citet{liu2022discontinuity} introduce an additional loss term that is a function of flow gradient, such that highly-compressed regions have a smaller weighting during loss evaluation. The methodology is used to simulate the 2D Riemann problem as governed by the compressible Euler equations. Their method improves performance of PDE-based losses and yields accuracy comparable with high-order numerical methods. 

Further feasible ideas for ML modelling of conservative hyperbolic systems can be drawn from traditional numerical techniques. As such we briefly review these techniques\footnote{If the reader is already familiar with finite volume methods and specifically Godunov methods, they may skip the next two paragraphs.}. The first major advancement in finite volume methods (FVMs) for compressible flows and hyperbolic systems can be traced back to the work of Godunov in 1959 \cite{godunov1959finite}. The so-called Godunov scheme assumes the flow-field is piecewise constant and relies on the exact solution of the local Riemann problem at each cell interface to estimate the intercell fluxes \cite{toro2013riemann}. Godunov's scheme preserves monotonicity and is entropy-satisfying \cite{sweby2001godunov}. However, Godunov's theorem states that monotonicity-preserving constant-coefficient schemes can be at most 1st-order accurate \cite{godunov1959finite}. Since the Riemann problem has to be exactly solved at every interface and at every time step, the Godunov scheme is computationally expensive when applied to non-linear systems of conservation laws \cite{sweby2001godunov}. Nevertheless the work of Godunov inspired the development of numerous extensions to his scheme. These extensions focus on reducing computational burden by approximating the Riemann problem or the flux computation. Such schemes are known as approximate Riemann solvers. One such scheme developed by \citet{roe1981approximate} linearizes the quasi-linear form of hyperbolic conservative equations by replacing Jacobians with interval-wise constant matrices such that hyperbolicity, consistency, and conservation are still satisfied. The approximate Riemann problem is linear meaning solutions only admit discontinuities and not expansion fans. Roe's method is therefore not entropy-satisfying although entropy fixes have been proposed \cite{pelanti2001review}. Osher's method is similar to Roe's method but approximates the Riemann problem using simple waves, rather than discontinuities \cite{engquist1981one,osher1982upwind}, and is entropy-satisfying \cite{sweby2001godunov}. The methods of Roe and Osher assume $n$ intermediate states for $n$ conservation laws when approximating the Riemann problem. \citet{harten1983upstream} propose a simplification by assuming a Riemann solution of three states separated by two waves, which correspond to the fastest and slowest signal speeds emerging from the discontinuity at a cell interface. They name their algorithm the Harten, Lax, Van Leer solver (HLL). \citet{toro1994restoration} modify the HLL solver by assuming an additional middle wave that accounts for contact and shear waves, known as the HLLC solver. Both HLL and HLLC are entropy-satisfying \cite{sweby2001godunov}.

Godunov's theorem led to much research on developing higher-order non-linear schemes applicable to hyperbolic systems. \citet{harten1997high} postulates that monotonicity-preserving higher-order schemes are achievable provided the discrete total variation (TV) of the solution does not increase over time. He coins schemes that satisfy this non-increase in TV as total variation diminishing (TVD). The monotonic upstream-centered scheme for conservation laws (MUSCL) reconstructs piece-wise constant data into piecewise linear data to achieve 2nd-order accuracy \cite{van1977towards,van1979towards}. The reconstruction includes slope-limiting techniques to ensure the scheme is TVD. Similarly the piecewise parabolic method (PPM) is of 3rd-order accuracy \cite{colella1984piecewise,woodward1984numerical}. Essentially non-oscillatory schemes (ENO) involve a recursive piecewise polynomial reconstruction to the desired order of accuracy by iteratively including substencils for interpolation \cite{harten1986some, harten1987uniformly, shu1988efficient, shu1989efficient, harten1997uniformly}. The choice between candidate substencils is based on which provides the smoothest interpolation as indicated by smoothness indicators. ENO is not strictly TVD although oscillations tend to disappear if the solution is adequately resolved \cite{sweby2001godunov}. WENO is a TVD-extension of ENO that uses a weighted convex combination of all possible candidate substencils for reconstruction \cite{liu1994weighted}. Both ENO and WENO can achieve up to 5th-order accuracy. For a more detailed discussion of FVMs for hyperbolic systems we refer the reader to \cite{leveque2002finite, toro2013riemann}.

ML research that draws from traditional numerical techniques is as follows. \citet{patel2022thermodynamically} use a control volume PINN (cvPINN) which evaluates the physics-informed loss in integral form rather than PDE form. They use artificial viscosity and an entropy inequality penalization term to yield entropy-satisfying solutions. A TV penalization term is also used to prevent artificial oscillations developing in the solution, in a manner similar to \cite{tezaur2017advanced}. Their method proves to be superior to standard PDE-based PINNs when modelling 1D Euler and Bucky-Leverett equations. The use of three additional terms however leads to additional hyperparameter tuning. It is also unclear how the intercell flux is calculated in evaluating the control volume loss. \citet{hansen2023learning} take a probabilistic approach of combining the integral form of conservation laws with a Bayesian update to learn solutions of the generalized porous medium equation. \citet{bezgin2023jax} use a neural network to optimize the artificial viscosity parameter in the Lax-Friedrichs flux to yield a stable and less diffusive solution for the 2D Riemann problem. \citet{xiong2020roenets} use convolutional neural networks (CNNs) to predict the linearization matrices of the Roe solver from the flow solution to perform time-stepping. Their method improves over a traditional Roe solver when applied on the shock-tube problem. \citet{kossaczka2021enhanced, kossaczka2023deep} use neural networks to optimize the smoothness indicators of WENO schemes. They also add a regularization term to a MSE loss to penalize any overflows of the solution, thereby preventing artificial oscillations. Their method improves over other WENO variants when tested on numerous hyperbolic problems. Other similar works using ML to predict WENO weights and smoothness indicators are \cite{wang2019learning, bezgin2022weno3, li2022six}. \citet{wang2023fluxnet} make use of Rankine-Hugoniot (R-H) jump conditions to derive a regularizing term. The R-H conditions describe how physical quantities change across a shock wave as derived from conservation laws, eliminating entropy violations. They apply the methodology to simulate the 2D Riemann problem and yield higher accuracy in comparison to a numerical Roe solver. \citet{peyvan2024riemannonets} use neural operator learning to solve the 1D Riemann problem. Their method consists of networks for encoding spatial information and input (left) pressure to predict state variables at the final time. 

In this paper, we present the first use of Godunov losses for modelling of 2D hyperbolic systems. We define Godunov losses as FVM-based loss functions that fundamentally incorporate the approximate solution of the Riemann problem to estimate the intercell fluxes required in the control volume loss. In using such loss functions we hope to a) improve accuracy compared to regularized PDE-based PINN losses as well as non-Godunov FVM-based PINN losses and b) encourage entropy-satisfying non-oscillatory solutions with minimal regularization terms (that otherwise require tuning). Our approach is fully unsupervised. We examine performance of this Godunov loss function on forward and inverse problems pertaining to the 2D Riemann problem. The 2D Riemann problem gives rise to varying complex flow patterns for different initial conditions (or configurations) and thus serves as a common canonical problem for testing compressible flow solvers. For the forward problem we time-step six different configurations of the Riemann problem using Conv-LSTM models. For the inverse problem we focus on one configuration and perform super-resolution reconstruction at three different input resolutions. The super-resolution is performed via a very deep super-resolution (VDSR) model \cite{kim2016accurate}. Results indicate superiority of the Godunov loss compared to baseline physical losses (PDE-based and non-Godunov FVM-based) in terms of accuracy and convergence.


The paper is structured as follows: in Section \ref{sec:back} we outline the equations, the ML architectures, and the formulation of the Godunov loss. In Section \ref{sec:exp} we describe the experimental setups and discuss results. Finally we make concluding remarks in Section \ref{sec:conc}.
\newcommand*{\Scale}[2][4]{\scalebox{#1}{$#2$}}%
\newcommand*{\Resize}[2]{\resizebox{#1}{!}{$#2$}}%

\section{Methodology}\label{sec:back}

\subsection{Equations}

We are interested in conservative hyperbolic PDEs of the general form:

\begin{equation}\label{eq:gencons}
    \partial_{t}\mathbf{Q} + \partial_{x}\mathbf{F}\left(\mathbf{Q}\right) + \partial_{y}\mathbf{G}\left(\mathbf{Q}\right) + \partial_{z}\mathbf{H}\left(\mathbf{Q}\right) = \mathbf{0},
\end{equation}

with associated boundary conditions (BCs), where $\mathbf{Q}$ is the vector of conserved quantities and $\mathbf{F}$, $\mathbf{G}$ and $\mathbf{H}$ are the flux vectors in the $x$, $y$ and $z$ directions. Re-writing equation \eqref{eq:gencons} in quasilinear form:

\begin{equation}
    \partial_{t}\mathbf{Q} + \partial_{\mathbf{Q}}\mathbf{F}\partial\mathbf{Q}_{x} + 
    \partial_{\mathbf{Q}}\mathbf{G}\partial\mathbf{Q}_{y} +
    \partial_{\mathbf{Q}}\mathbf{H}\partial\mathbf{Q}_{z} = \mathbf{0},
\end{equation}

allows us to define a hyperbolic system.

\textbf{Definition} \cite{toro2013riemann}: \textit{A system is said to be hyperbolic if the eigenvalues of Jacobians $\partial_{\mathbf{Q}}\mathbf{F}$, $\partial_{\mathbf{Q}}\mathbf{G}$, and $\partial_{\mathbf{Q}}\mathbf{H}$ are all real with a corresponding set of linearly independent eigenvectors. The system is said to be strictly hyperbolic if the eigenvalues are distinct.}

In this paper we focus on the 2D Euler equations:

\begin{equation}
    {\partial_t}
\begin{pmatrix}
        \rho      \\
        \rho u      \\
        \rho v      \\
         E      \\
\end{pmatrix} + {\partial_x}
\begin{pmatrix}
        \rho u      \\
        \rho u^{2} + p \\
        \rho u v \\
        u\left(E + p\right) \\
\end{pmatrix} + {\partial_y}
\begin{pmatrix}
        \rho v      \\
        \rho u v    \\
        \rho v^{2} + p \\
        v\left(E + p\right) \\
    \end{pmatrix} = \mathbf{0}, \\
\end{equation}

with associated Neumann BCs $\partial_{\mathbf{n}} \mathbf{Q} = \mathbf{0}$. To close the system energy $E$ is expressed in terms of velocities, pressure, density and ratio of specific heats $\gamma$:

\begin{equation}
        E := \frac{1}{2}\rho\left(u^2 + v^2\right) + \frac{p}{\left(\gamma - 1\right)}.
\end{equation}

\subsection{Architectures}

\begin{figure*}[h]
\begin{center}
\centerline{\includegraphics[width=1.0\textwidth]{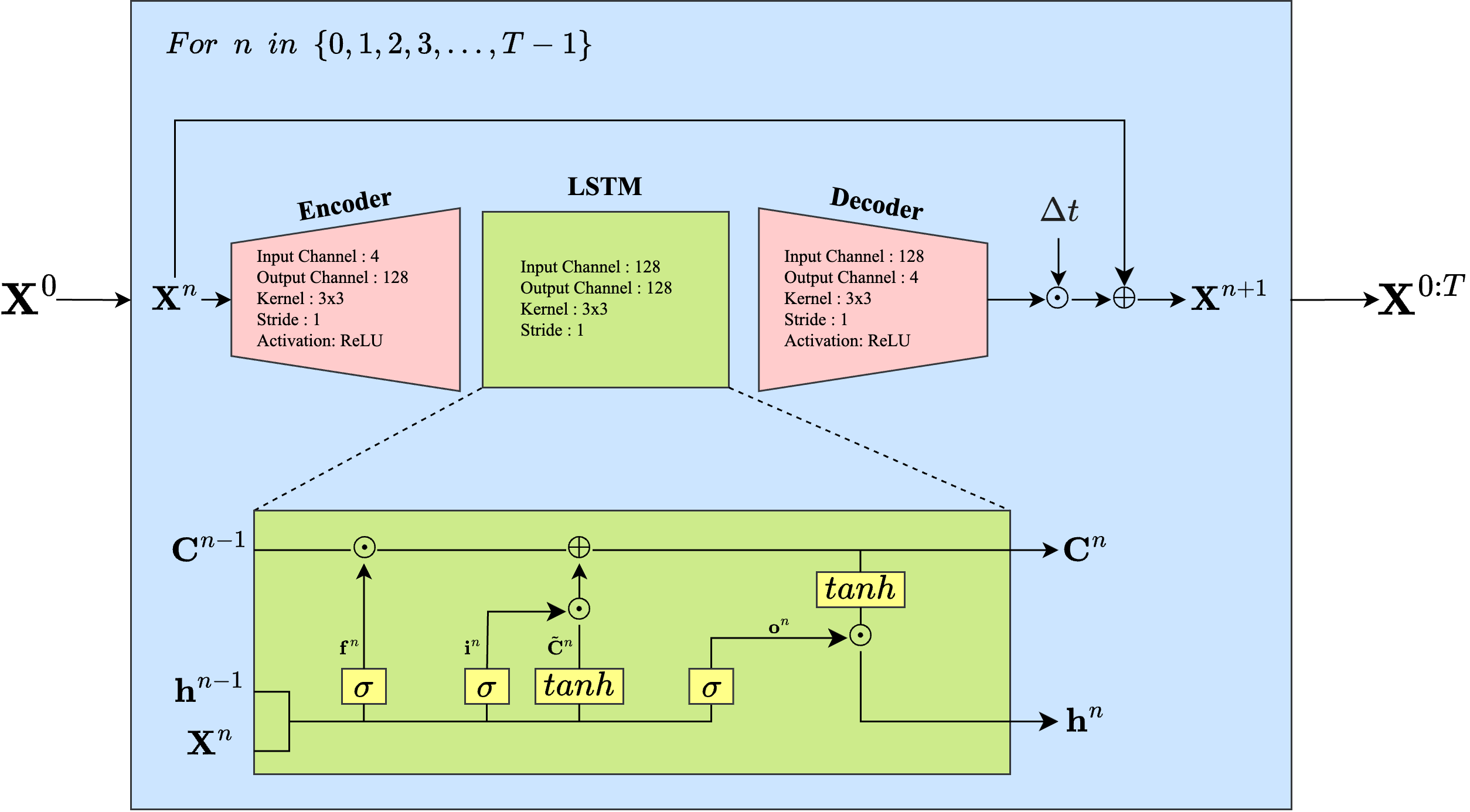}}
\caption{The Conv-LSTM architecture, which takes as input the initial  condition $\mathbf{X}^0$ and iteratively updates it $T-1$ times, storing each update in a list such that the final output is a sequence of snapshots $\mathbf{X}^{0:T}$. The hidden and cell states when $n=0$, i.e., $\mathbf{h}^{i}$ and $\mathbf{C}^{i}$, are initialized randomly. All linear operations in the encoder, LSTM module and decoder are convolutional.}
\label{fig:timestep}
\end{center}
\vskip -0.3in
\end{figure*}

For the forward time-stepping problem, we implement a Conv-LSTM architecture with a global residual connection to mimic the forward Eulerian scheme:

\begin{equation}\label{eq:convLSTM}
    \mathbf{X}^{n+1} = \mathbf{X}^n + \Delta \mathbf{X}^n = \mathbf{X}^n + \Delta t\cdot\bm{\mathcal{F}}_{TS}\left(\mathbf{X}^n; \mathbf{\Theta}_{TS}\right),
\end{equation}

where $\mathbf{X}$ is the feature, $\bm{\mathcal{F}}_{TS}$ is the time-stepper network parameterized by weights $\mathbf{\Theta}_{TS}$, $\Delta t$ is the step size, and $n$ is the temporal index. The residual connection in equation \eqref{eq:convLSTM} automatically imposes the initial condition of the PDEs of concern. The network $\bm{\mathcal{F}}_{TS}$ comprises an encoder-LSTM-decoder combination ($\bm{\mathcal{F}}_{Enc}$, $\bm{\mathcal{F}}_{LSTM}$, $\bm{\mathcal{F}}_{Dec}$):

\begin{align}
    &\Bar{\mathbf{X}}^{n} = \bm{\mathcal{F}}_{Enc}\left(\mathbf{X}^n; \mathbf{\Theta}_{Enc}\right) \\
    &\mathbf{h}^n = \bm{\mathcal{F}}_{LSTM}\left(\Bar{\mathbf{X}}^{n}, \mathbf{h}^{n-1}, \mathbf{C}^{n-1}; \mathbf{\Theta}_{LSTM}\right)\\
    &\Delta \mathbf{X}^n = \Delta t \cdot \bm{\mathcal{F}}_{Dec}\left({\mathbf{h}^n}; \mathbf{\Theta}_{Dec}\right)
\end{align}

where $\Bar{\mathbf{X}}$ is the encoding of ${\mathbf{X}}$, $\mathbf{h}$ is the hidden state and $\mathbf{C}$ is the cell state. The LSTM module $\bm{\mathcal{F}}_{LSTM}$ comprises an input gate $\mathbf{i}$, forget gate $\mathbf{f}$, and output gate $\mathbf{o}$, parameterized by weights \{$\mathbf{W}_i, \mathbf{W}_f,\mathbf{W}_c, \mathbf{W}_o$\} and biases \{$\mathbf{b}_i, \mathbf{b}_f, \mathbf{b}_c, \mathbf{b}_o$\}. Equations \eqref{eq:LSTMi} - \eqref{eq:LSTMf} outline the LSTM operations \cite{hochreiter1997long}:

\begin{align}
    &\mathbf{i}^{n} = \sigma \left( \mathbf{W}_i * [\Bar{\mathbf{X}}^{n}, \mathbf{h}^{n-1}] + \mathbf{b}_i \right) \label{eq:LSTMi} \\ 
    &\mathbf{f}^{n} = \sigma \left( \mathbf{W}_f * [\Bar{\mathbf{X}}^{n}, \mathbf{h}^{n-1}] + \mathbf{b}_f \right) \\
    &\Tilde{\mathbf{C}}^{n-1} = \text{tanh} \left( \mathbf{W}_c * [\Bar{\mathbf{X}}^{n}, \mathbf{h}^{n-1}] + \mathbf{b}_c \right) \\
    &\mathbf{C}^{n} = \mathbf{f}^{n} \odot \mathbf{C}^{n-1} +  \mathbf{i}^{n} \odot \Tilde{\mathbf{C}}^{n-1} \\ 
    &\mathbf{o}^{n} = \sigma \left( \mathbf{W}_o * [\Bar{\mathbf{X}}^{n}, \mathbf{h}^{n-1}] + \mathbf{b}_o \right) \\
    &{\mathbf{h}}^{n} = \mathbf{o}^{n} \odot \text{tanh}\left(\mathbf{C}^{n}\right) \label{eq:LSTMf}
\end{align}

where $\sigma$ is the sigmoid function, $*$ is the convolutional operation, and $\odot$ is the Hadamard product. Figure \ref{fig:timestep} illustrates the time-stepping framework.

For the inverse super-resolution problem, we first pass the low resolution input $\mathbf{X}_{LR}$ into a series of convolutional and bilinear upsampling layers $\bm{\mathcal{F}}_{US}$ to get to the desired resolution, and then pass this output to a VDSR-based module $\bm{\mathcal{F}}_{VDSR}$:

\begin{equation}\label{eq:vdsr}
    \mathbf{X}_{HR} = \bm{\mathcal{F}}_{VDSR}\left( \bm{\mathcal{F}}_{US}\left( \mathbf{X}_{LR}; \mathbf{\Theta}_{US}\right); \mathbf{\Theta}_{VDSR}\right).
\end{equation}

\begin{figure*}[h]
\begin{center}
\centerline{\includegraphics[width=1.0\textwidth]{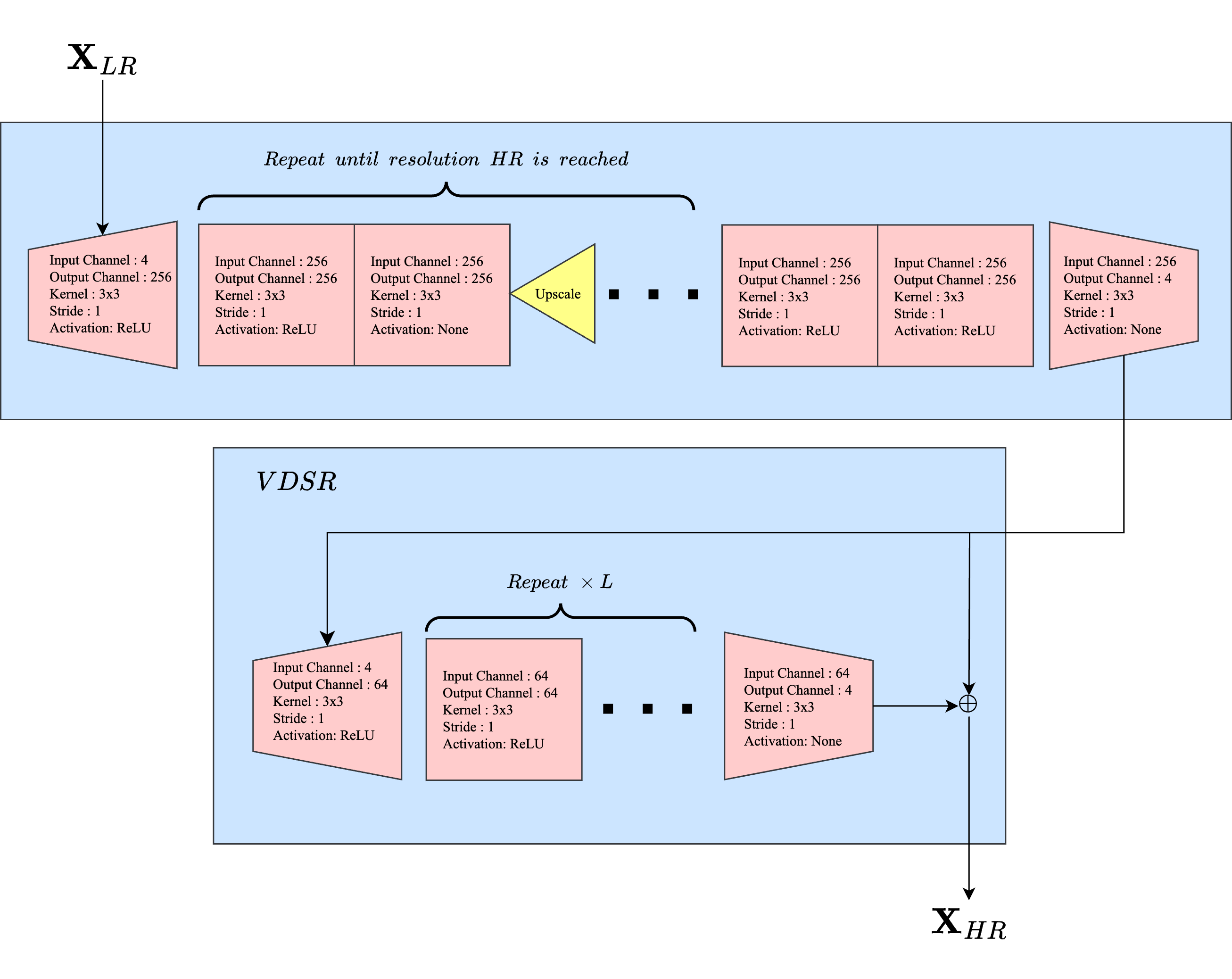}}
\caption{The super-resolution architecture. Input $\mathbf{X}_{LR}$ first passes through a series of convolutional and upscaling layers to get to the desired resolution. Upscaling is done via bilinear interpolation, with a scale factor of 2. The output from these layers is then passed to the VDSR network which is essentially a very deep convolutional neural network (number of hidden layers $L\sim20$ typically) with a residual connection. The residual connection allows for residual learning, which predicts the difference between the input and the desired output, enabling focus on fine details of the flow/image.}
\label{fig:superresdia}
\end{center}
\vskip -0.3in
\end{figure*}

The purpose of VDSR is to sharpen an image (without increasing resolution) by making use of a very deep convolutional network that is stabilized by a residual connection from its input. See \cite{kim2016accurate} for additional details about VDSR. Figure \ref{fig:superresdia} illustrates the reconstruction process.

A plausible choice for feature $\mathbf{X}$ is the set of primitives $\left( \rho, u, v, E \right)^T$. However we choose to work with:

\begin{equation}
    \mathbf{X} = \left( \rho, u, v, E - \frac{1}{2}\rho(u^{2} + v^{2}) \right)^T,
\end{equation}

and apply the following transformation on the final output from the network:

\begin{equation}
    \left( \rho, u, v, E - \frac{1}{2}\rho\left(u^{2} + v^{2}\right) \right)^T \longrightarrow  \left( \left|{\rho}\right|, u, v, \left|E - \frac{1}{2}\rho\left(u^{2} + v^{2}\right)\right|\right)^T,
\end{equation}

such that we ensure positivity conservation for the set of all states as a hard constraint \cite{batten1997choice}:

\begin{equation}
    \left\{ \left( \rho, u, v, E \right)^T, \,\,\,\, \rho > 0, \,\,\,\,  E - \frac{1}{2}\rho\left(u^{2} + v^{2}\right) > 0 \right\}.
\end{equation}

\begin{figure*}[h]
\begin{center}
\centerline{\includegraphics[width=0.50\textwidth]{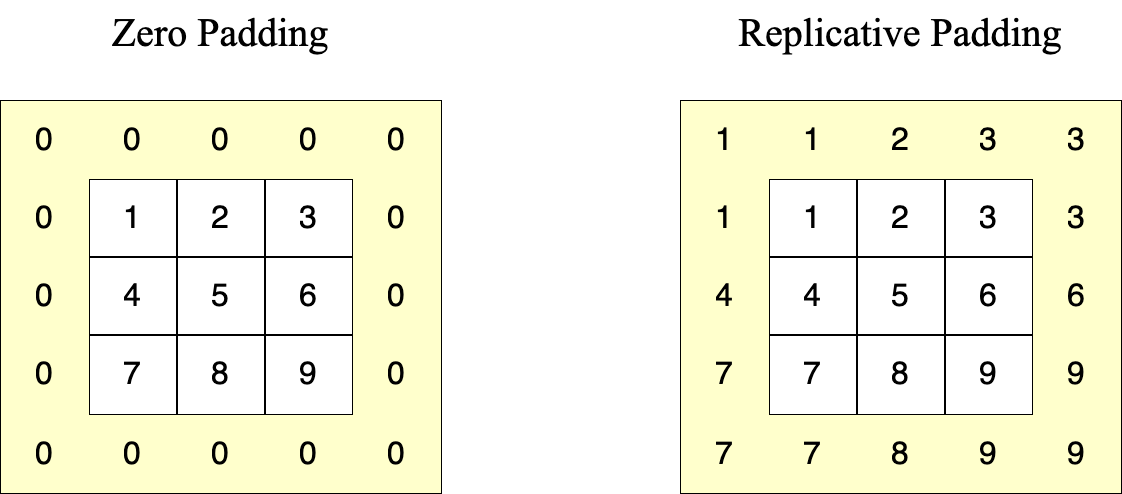}}
\caption{Illustration of replicative padding as opposed to the default (zero) padding used in ML libraries.}
\label{fig:padding}
\end{center}
\vskip -0.3in
\end{figure*}

Additionally, constant Neumann BCs are embedded into these convolutional architectures by padding with values interpolated from the interior domain. The simplest case, where the BC is $\partial_{\mathbf{n}} \mathbf{Q} = \mathbf{0}$, is incorporated using replicative padding on the features, as shown in Figure \ref{fig:padding}.

\subsection{Godunov Loss}

A PINN typically evaluates its loss by directly evaluating derivatives in the relevant PDEs. However the differential form of the governing equations is invalid in regions of shocks that may manifest in hyperbolic systems. An integral form on the other hand is applicable across shocks. The approach presented here for evaluating the loss function is finite volume-based and inherently aims to achieve three desirable properties. Firstly, the approach aims to satisfy the weak form of the conservation law. In 1D, over domain $\left[t_{1}, t_{2}\right] \times \left[x_{1}, x_{2}\right]$, the weak conservation law is:

\begin{equation}\label{eq:integ}
    \int_{x_{1}}^{x_{2}} \left[ \mathbf{Q}\left(x,t_{2}\right) - \mathbf{Q}\left(x,t_{1}\right)\right]dx  + \int_{t_1}^{t_2}\left[\mathbf{F}\left(\mathbf{Q}\left(x_{2},t\right)\right) -\mathbf{F}\left(\mathbf{Q}\left(x_{1},t\right)\right)\right]dt = \mathbf{0}.
\end{equation}

Secondly, the approach aims to satisfy the weak form of the entropy condition to encourage physically correct shocks. In 1D, over the same domain, the condition is:

\begin{equation}\label{eq:integent}
    \int_{x_{1}}^{x_{2}} \left[\Phi\left(\mathbf{Q}\left(x,t_{2}\right)\right) -\Phi\left(\mathbf{Q}\left(x,t_{1}\right)\right)\right]dx  + \int_{t_1}^{t_2}\left[\mathbf{\Tilde{F}}\left(\mathbf{Q}\left(x_{2},t \right) \right) - \mathbf{\Tilde{F}}\left(\mathbf{Q}\left(x_{1},t\right)\right)\right]dt \leq 0,
\end{equation}

 where convex entropy function $\Phi\left(\mathbf{Q}\right)$ satisfies $\partial_{\mathbf{Q}}\Phi \partial_{\mathbf{Q}}\mathbf{F} = \partial_{\mathbf{Q}}\mathbf{\Tilde{F}}$, where $\mathbf{\Tilde{F}}$ is the entropy flux. Thirdly, the approach aims to preserve solution monotonicity (i.e., prevent artificial oscillations): see Section 1 in \citet{harten1983upstream} for a detailed discussion of these three properties. A class of FVMs possessing these properties are known as Godunov methods. In the following subsections we outline the Godunov method then derive our so-called Godunov loss from it. Godunov's method is outlined in 1D for brevity but is then extended to 2D when formulating the loss, as we are ultimately interested in modelling the 2D Euler equations.

\subsubsection{Godunov Methods}

Godunov methods are based on solving local Riemann problems at each FVM cell interface. Assuming piecewise-constant $\mathbf{Q}\left(x_{i}, t^{n}\right) = \mathbf{Q}_{i}^n$, the local Riemann problem at cell interface $i + 1/2$, at the $n$th time:  

\begin{equation}\label{eq:riem}
\mathbf{Q}\left(\Bar{x},\Bar{t} = 0\right) =\begin{cases}
    \mathbf{Q}_l \equiv \mathbf{Q}_i^n & \Bar{x}<0,\\
    \mathbf{Q}_r \equiv \mathbf{Q}_{i+1}^n & \Bar{x}>0,
  \end{cases}
\end{equation}

depends only on states $\mathbf{Q}_l$ and $\mathbf{Q}_r$, and ratio of local coordinates $\Bar{x}/\Bar{t}$, where $\Bar{x} = x - x_{i+\frac{1}{2}}$ and $\Bar{t} = t - t^n$. Since signals travel with finite velocity bounded by a minimum $a_l$ and a maximum $a_r$, then \cite{harten1983upstream}:

\begin{equation}\label{eq:finitespeeds}
    \mathbf{Q}\left(\Bar{x}/\Bar{t};\mathbf{Q}_{l}, \mathbf{Q}_{r}\right) =\begin{cases}
    \mathbf{Q}_l \equiv \mathbf{Q}_i^n & \Bar{x}/\Bar{t} \leq a_l,\\
    \mathbf{Q}_r \equiv \mathbf{Q}_{i+1}^n & \Bar{x}/\Bar{t} \geq a_r.
  \end{cases} 
\end{equation}

By assuming $\lambda|a_{max}| < 1/2$, where $a_{max}$ is the largest signal speed in the domain and $\lambda = \Delta t / \Delta x$, then by equation \eqref{eq:finitespeeds} there is no interaction between neighbouring Riemann problems and the solution $\mathbf{\hat{Q}}$ in the interval $\left[t^{n}, t^{n+1}\right] \times \left[x_{i}, x_{i+1}\right]$ can be expressed exactly in terms of the solution to the local Riemann problem \cite{toro2013riemann}:

\begin{equation}\label{eq:locrie}
    \mathbf{\hat{Q}}\left(x,t\right) = \mathbf{Q}\left(\Bar{x}/\Bar{t}; \mathbf{Q}_{i}^{n},\mathbf{Q}_{i+1}^{n}\right).
\end{equation}

A piecewise-constant approximation at the next time $t^{n+1}$ is then given by:

\begin{equation}\label{eq:aver}
    \mathbf{Q}_{i}^{n+1} = \frac{1}{\Delta x}\int_{x_{i-\frac{1}{2}}}^{x_{i + \frac{1}{2}}} \mathbf{\hat{Q}}\left(x, t^{n} + \Delta t\right)dx,
\end{equation}

or in terms of local Riemann problems as:

\begin{equation}
    \mathbf{Q}_{i}^{n+1} 
    = \frac{1}{\Delta x}\int_{0}^{\frac{\Delta x}{2}}\mathbf{Q}\left(x/\Delta t; \mathbf{Q}_{i-1}^n, \mathbf{Q}_{i}^{n}\right)dx
    + \frac{1}{\Delta x}\int_{-\frac{\Delta x}{2}}^{0}\mathbf{Q}\left(x/\Delta t; \mathbf{Q}_{i}^{n}, \mathbf{Q}_{i+1}^n\right)dx.
\end{equation}

If we relax the time-step criterion such that $\lambda|a_{max}| \leq 1$, then waves issuing from interface $x_{i - \frac{1}{2}}$ do not reach  $x_{i + \frac{1}{2}}$ in the interval $\left[t^{n}, t^{n+1}\right]$, and vice versa. It can then be shown that, by applying equation \eqref{eq:integ} over $\left[t^{n}, t^{n+1}\right] \times \left[x_{i-\frac{1}{2}}, x_{i+\frac{1}{2}}\right]$, we can arrive at a conservative expression at the next time level \cite{harten1983upstream}:

\begin{equation}\label{eq:consform}
    \mathbf{Q}_{i}^{n+1} = \mathbf{Q}_{i}^{n} - \lambda\left[\mathbf{F\left(\bm{\mathfrak{R}}\right)}_{i + \frac{1}{2}}^{n} - \mathbf{F\left(\bm{\mathfrak{R}}\right)}_{i - \frac{1}{2}}^{n}\right],
\end{equation}

\begin{equation}\label{eq:fhow}
    \mathbf{F\left(\bm{\mathfrak{R}}\right)}_{i + \frac{1}{2}}^{n} = \mathbf{F}\left(\mathbf{Q}\left(\Bar{x}/\Bar{t}=0;\mathbf{Q}_{i}^{n},\mathbf{Q}_{i+1}^{n}\right)\right), \,\,\,\,\,\, 
    \mathbf{F\left(\bm{\mathfrak{R}}\right)}_{i - \frac{1}{2}}^{n} =  \mathbf{F}\left(\mathbf{Q}\left(\Bar{x}/\Bar{t}=0;\mathbf{Q}_{i-1}^{n},\mathbf{Q}_{i}^{n}\right)\right).
\end{equation}

For proof, see Chapter 6 in \citet{toro2013riemann}. $\mathbf{\hat{Q}}$ satisfies the weak form of the entropy condition (inequality \eqref{eq:integent}):

\begin{equation}\label{eq:weakent}
    \int_{x_{i-\frac{1}{2}}}^{x_{i + \frac{1}{2}}} \Phi\left(\mathbf{\hat{Q}}\left(x, t^{n} + \Delta t\right)\right)dx 
    \leq \Delta\Phi\left(\mathbf{Q}_{i}^n\right) - \Delta t \left[ \mathbf{\Tilde{F}\left(\bm{\mathfrak{R}}\right)}_{i+\frac{1}{2}}^{n} - \mathbf{\Tilde{F}\left(\bm{\mathfrak{R}}\right)}_{i-\frac{1}{2}}^{n}\right],
\end{equation}

\begin{equation}
    \mathbf{\Tilde{F}\left(\bm{\mathfrak{R}}\right)}_{i + \frac{1}{2}}^{n} = \mathbf{\Tilde{F}}\left(\mathbf{Q}\left(\Bar{x}/\Bar{t}=0;\mathbf{Q}_{i}^{n},\mathbf{Q}_{i+1}^{n}\right)\right), \,\,\,\,\,\,
    \mathbf{\Tilde{F}\left(\bm{\mathfrak{R}}\right)}_{i - \frac{1}{2}}^{n} =  \mathbf{\Tilde{F}}\left(\mathbf{Q}\left(\Bar{x}/\Bar{t}=0;\mathbf{Q}_{i-1}^{n},\mathbf{Q}_{i}^{n}\right)\right).
\end{equation}

Since $\Phi$ is a convex function, we can use Jensen's inequality along with inequality \eqref{eq:weakent} to deduce that Godunov's method satisfies the entropy inequality \cite{harten1983upstream}.

Due to averaging in equation \eqref{eq:aver}, information contained in the exact solution of the Riemann problem is lost. This implies more efficient Godunov methods are achievable by assuming a simpler structure to the Riemann solution a-priori. These so-called Godunov-type methods are also known as approximate Riemann solvers. Like Godunov schemes, it can be shown that Godunov-type schemes also satisfy the entropy condition and can be written in conservation form to arrive at equation \eqref{eq:consform}, see Section 3 in \citet{harten1983upstream}.

\subsubsection{Construction of Loss}\label{sec:loss}

We are interested in 2D hyperbolic problems, and therefore extend equation \eqref{eq:consform} to 2D:

\begin{equation}\label{eq:consform2D}
    \mathbf{Q}_{i,j}^{n+1} = \mathbf{Q}_{i,j}^{n} - \lambda_{x}\left[{\mathbf{F}\left(\bm{\mathfrak{R}}\right)}_{i + \frac{1}{2},j}^{n} - {\mathbf{F}\left(\bm{\mathfrak{R}}\right)}_{i - \frac{1}{2},j}^{n}\right] 
    - \lambda_{y}\left[{\mathbf{G}\left(\bm{\mathfrak{R}}\right)}_{i,j + \frac{1}{2}}^{n} - {\mathbf{G}\left(\bm{\mathfrak{R}}\right)}_{i,j - \frac{1}{2}}^{n}\right],
\end{equation}

where by analogy to equation \eqref{eq:fhow} we can express fluxes $\mathbf{G}$ as:

\begin{equation}\label{eq:ghow}
    \mathbf{G\left(\bm{\mathfrak{R}}\right)}_{j + \frac{1}{2}}^{n} = \mathbf{G}\left(\mathbf{Q}\left(\Bar{y}/\Bar{t}=0;\mathbf{Q}_{j}^{n},\mathbf{Q}_{j+1}^{n}\right)\right), \,\,\,\,\,\, 
    \mathbf{G\left(\bm{\mathfrak{R}}\right)}_{j - \frac{1}{2}}^{n} =  \mathbf{G}\left(\mathbf{Q}\left(\Bar{y}/\Bar{t}=0;\mathbf{Q}_{j-1}^{n},\mathbf{Q}_{j}^{n}\right)\right).
\end{equation}

For compactness we re-write equation \eqref{eq:consform2D} as:

\begin{equation}\label{eq:consform2dcom}
    \mathbf{Q}_{i,j}^{n+1} = \mathbf{Q}_{i,j}^{n} - \lambda_{x}\left[\Delta{\mathbf{F}\left(\bm{\mathfrak{R}}\right)}_{i,j}^{n} \right]
    - \lambda_{y}\left[\Delta{\mathbf{G}\left(\bm{\mathfrak{R}}\right)}_{i,j}^{n}\right],
\end{equation}

\begin{equation}
    \Delta{\mathbf{F}\left(\bm{\mathfrak{R}}\right)}_{i,j}^{n} = {\mathbf{F}\left(\bm{\mathfrak{R}}\right)}_{i + \frac{1}{2},j}^{n} - {\mathbf{F}\left(\bm{\mathfrak{R}}\right)}_{i - \frac{1}{2},j}^{n}, \,\,\,\,\,\,
    \Delta{\mathbf{G}\left(\bm{\mathfrak{R}}\right)}_{i,j}^{n} = {\mathbf{G}\left(\bm{\mathfrak{R}}\right)}_{i,j + \frac{1}{2}}^{n} - {\mathbf{G}\left(\bm{\mathfrak{R}}\right)}_{i,j - \frac{1}{2}}^{n},
\end{equation}

where $\lambda_{x} = \Delta t / \Delta x$ and $\lambda_{y} = \Delta t / \Delta y$. Re-arranging equation \eqref{eq:consform2dcom} we can express the Godunov loss as:

\begin{equation}\label{eq:godloss}
    \mathcal{L}_{G} :=  \frac{\bm{\omega}}{N} \cdot \mathop{\Scale[2]{\sum}}_{i,j,n} \left( 
    \mathbf{Q}_{i,j}^{n+1} - \mathbf{Q}_{i,j}^{n}  + \lambda_{x}\left[\Delta{\mathbf{F}\left(\bm{\mathfrak{R}}\right)}_{i,j}^{n} \right] 
    + \lambda_{y}\left[\Delta{\mathbf{G}\left(\bm{\mathfrak{R}}\right)}_{i,j}^{n}\right]\right)^{2},
\end{equation}

where $N$ is total number of points in space and time, and $\bm{\omega}$ weighs the residual contribution from each equation in a PDE system.

Riemann solvers mainly differ by how the fluxes $\mathbf{F}\left(\bm{\mathfrak{R}}\right)$ and $\mathbf{G}\left(\bm{\mathfrak{R}}\right)$ are estimated. To demonstrate the effectiveness of Godunov losses in our experiments, we use as an example the HLLC method to determine the intercell flux from  the ML prediction. However, we emphasize that any Riemann flux estimator could be used instead. 

\begin{figure}
\vskip 0.2in
\begin{center}
\centerline{\includegraphics[width=0.4\textwidth]{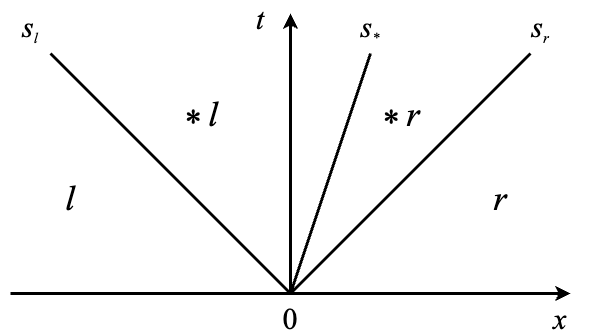}}
\caption{The three-wave model assumed by the HLLC solver. The left and right characteristic lines correspond to the fastest and slowest signals, $s_l$ and $s_r$, emerging from the cell interface at $x=0$. The middle characteristic corresponds to the wave of speed $s_*$ which accounts for contact and shear waves.}
\label{fig:hllc}
\end{center}
\vskip -0.3in
\end{figure}

An outline of the HLLC algorithm is now presented. The HLLC method assumes a three-wave structure to the Riemann problem at each cell interface, as shown in Figure \ref{fig:hllc}. In order to determine the HLLC flux normal to a cell interface for the 2D Euler equations, it is sufficient to consider a $x$-split version of the 2D Euler equations because of rotational invariance \cite{toro2019hllc}:

\begin{equation}
    {\partial_t}
\begin{pmatrix}
        \rho      \\
        \rho u      \\
        \rho v      \\
         E      \\
\end{pmatrix} + {\partial_x}
\begin{pmatrix}
        \rho u      \\
        \rho u^{2} + p \\
        \rho u v \\
        u( E + p) \\
\end{pmatrix} = \mathbf{0} 
\end{equation}

Given this, we then estimate the pressure in the star region $\Bar{p}_{*}$. A reliable approximation is \cite{toro2019hllc}:

\begin{equation}
    \Bar{p}_{*} = \left[ \frac{a_l + a_r - \frac{\gamma -1}{2}(u_r - u_l)}{\frac{a_l}{p_l^z} + \frac{a_r}{p_r^z}}\right]^{1/z}
\end{equation}

where $a = \sqrt{\gamma p / \rho}$ is the speed of sound and $z=\left(\gamma - 1\right)/2\gamma$. Wave speeds $s_l$ and $s_r$ are then estimated as \cite{davis1988simplified, toro2019hllc}:

\begin{equation}
    s_l= u_l - a_{l}q_{l}, \;\;\;\; s_r = u_r + a_{r}q_{r}, \;\;\;\;\;\;
    q_{k} = \begin{cases}
    1 & \text{if $\Bar{p}_{*} \leq p_k$},\\
    \left[ 1 + (\frac{\gamma + 1}{2\gamma})(\frac{\Bar{p}_{*}}{p_k} - 1 )\right]^{\frac{1}{2}} & \text{if $\Bar{p}_{*} > p_k$}
  \end{cases}
\end{equation}

where $k = \{l,r\}$. We then use $s_l$ and $s_r$ to compute intermediate speed $s_*$ \cite{toro2019hllc}:

\begin{equation}
    s_{*} = \frac{p_r - p_l + \rho_{l}u_{l}(s_l - u_l) 
    - \rho_{r}u_{r}(s_r - u_r)}{ \rho_{l}(s_l - u_l) 
    - \rho_{r}(s_r - u_r)}.
\end{equation}

The intermediate conservative vectors $\mathbf{Q}_{*l}$ and $\mathbf{Q}_{*r}$ are then computed using $s_*$, $s_l$ and $s_r$ as \cite{toro2019hllc}:

\begin{equation}
    \mathbf{Q}_{*k} = \rho_{k}\left(\frac{s_k - u_k}{s_k - s_{*}}\right)\begin{pmatrix}
        1      \\
        s_{*}      \\
        v_k      \\
        \frac{E_k}{\rho_k} + (s_{*} - u_k)\left[s_{*} + \frac{p_k}{\rho_k(s_k - u_k)}\right]      
\end{pmatrix}.
\end{equation}

Finally, the intermediate flux vectors $\mathbf{F}_{*l}$ and $\mathbf{F}_{*r}$ are computed from the state vectors so that the flux $\mathbf{F}_{i + \frac{1}{2}}^n$ is computed using equation \eqref{eq:HLLC}.

\begin{equation}\label{eq:HLLC}
    \mathbf{F}_{*k} = \mathbf{F}_k + s_k(\mathbf{Q}_{*k} - \mathbf{Q}_k), \;\;\;\;\;\;
    \mathbf{F}_{i + \frac{1}{2}}^n = \begin{cases}
        \mathbf{F}_l & 0 \leq s_l,\\
        \mathbf{F}_{*l} & s_l \leq 0 \leq s_{*},\\
        \mathbf{F}_{*r} &  s_{*} \leq 0 \leq s_{r},\\
        \mathbf{F}_{r} &   0 \geq s_{r}.
    \end{cases}
\end{equation}

Due to rotational invariance, an analogous procedure allows us to determine the flux $\mathbf{G}_{j + \frac{1}{2}}^n$.

\section{Experiments}\label{sec:exp}

\subsection{Forward Problem: Time-Stepping}

We examine the performance of the proposed Godunov loss in time-stepping the 2D Riemann problem at training and inference time. The 2D Riemann problem is a canonical problem in compressible fluid dynamics. A unit square domain $\left[0,1\right] \times \left[0,1\right]$ is divided into four quadrants where each quadrant is uniformly initialized with its own set of $\left(\rho, u, v, E\right)^T$. The quadrants are numbered 1 to 4, anticlockwise from the top-right quadrant. The initialization of the quadrants determines the types of the four waves separating them. There are three wave types - rarefaction $R$, shock $S$, and contact $J$, characterized by their thermodynamic relations across the wave \cite{schulz1993classification}. Given these relations there exist nineteen physically admissible initial conditions representing different combinations of $R$, $J$, and $S$ \cite{schulz1993classification}. These nineteen configurations can be grouped into six classes characterized by numbers of each wave type:

\begin{equation*}
    4R \,\,\,\,\,\,\,\,\,\,\,\, 4S\,\,\,\,\,\,\,\,\,\,\,\, 4J \,\,\,\,\,\,\,\,\,\,\,\, 2R+2J\,\,\,\,\,\,\,\,\,\,\,\,2S+2J\,\,\,\,\,\,\,\,\,\,\,\,R+S+2J
\end{equation*}

For the purposes of this paper we examine a representative configuration from each class:

\begin{equation*}
R_{21}^{+}R_{32}^{+}R_{34}^{+}R_{41}^{+}\,\,\,\,\,\,\,\,\,S_{21}^{-}S_{32}^{+}S_{34}^{+}S_{41}^{-}\,\,\,\,\,\,\,\,\,J_{21}^{-}J_{32}^{+}J_{34}^{-}J_{41}^{+}\,\,\,\,\,\,\,\,
R_{21}^{-}J_{32}^{-}J_{34}^{-}R_{41}^{-}\,\,\,\,\,\,\,\,\,S_{21}^{+}J_{32}^{+}J_{34}^{+}S_{41}^{+}\,\,\,\,\,\,\,\,\, R_{21}^{-}J_{32}^{-}J_{34}^{+}S_{41}^{+}
\end{equation*}

where subscript $lr = \{21, 32, 34, 41\}$. The thermodynamic relations across each wave type and the initialization values are outlined in \ref{app:convalues}. For each configuration, we train a separate Conv-LSTM model (Figure \ref{fig:timestep}) to time-step the initial condition at $n=0$ to $n = 75$, and then examine inference performance (on the same configuration) from $n=0$ to $n = 150 $, where $n = t/\Delta t$ and $\Delta t = 0.002 s$ \footnote{Training and inference are performed on a single NVIDIA A100 GPU, requiring up to $\sim$ 20 Gb and $\sim$ 30 Gb of RAM, as well as $\sim$ 24 hrs and $\sim 1s$ of compute time, respectively.}. The model is Kaiming-initialized, and training is performed in a series of fine-tunes which train $5$ time-steps further than the previous fine-tune, using an Adam optimizer and a learning rate of $3 \times 10^{-5}$ throughout. We use a single convolutional-swish layer to project the $128^2$ input with 4 channels to a representation with 128 channels. The LSTM module then time-steps this representation before projecting back to 4 channels using a swish-deconvolutional layer. We use a kernel size of 3 for all layers \footnote{Tuning of hyperparameters is performed through a random grid-search of different combinations of hyperparameters.}. We implement this procedure with a Godunov loss $\mathcal{L}_G$ as described in equation \eqref{eq:godloss}. For comparisons we repeat the procedure using regularized PDE-based losses $\mathcal{L}_{Visc}$ and $\mathcal{L}_{TV+Ent}$, as well as a non-Godunov FVM-based loss $\mathcal{L}_{LF}$:

\begin{align}\label{eq:11}
&\mathcal{L}_{PDE} := \frac{\bm{\omega}}{N}\cdot\norm{\partial_{t}\mathbf{Q} + \partial_{x}\mathbf{F} + \partial_{y}\mathbf{G}}_{2}^{2}, \\[10pt]
\label{eq:22}
&\mathcal{L}_{Visc} := \frac{\bm{\omega}}{N}\cdot\norm{\partial_{t}\mathbf{Q} + \partial_{x}\mathbf{F} + \partial_{y}\mathbf{G}  - \alpha\left(\partial_{xx}{\mathbf{Q}} + \partial_{yy}{\mathbf{Q}}\right)}_{2}^{2}, \\[10pt]
\label{eq:33}
&\mathcal{L}_{TV+Ent} := \mathcal{L}_{PDE} + \frac{\beta_{1}}{N_{t}}\cdot\norm{\mathrm{max}\left({0}, \Delta TV\right)}_{2}^{2} + \frac{\beta_{2}}{N}\cdot\norm{\mathrm{max}\left(0, {\partial_t \Phi} + {\partial_{x} \mathbf{\Tilde{F}}} + {\partial_{y} \mathbf{\Tilde{G}}}\right)}_{2}^{2} , \\[10pt]
\label{eq:44}
&\mathcal{L}_{LF} := \frac{\bm{\omega}}{N} \cdot \mathop{\Scale[2]{\sum}}_{i,j,n} \left( 
    \mathbf{Q}_{i,j}^{n+1} - \mathbf{Q}_{i,j}^{n}  + \lambda_{x}\left[\Delta{\mathbf{F}\left(\cancel{\bm{\mathfrak{R}}}\right)}_{i,j}^{n} \right] 
    + \lambda_{y}\left[\Delta{\mathbf{G}\left(\cancel{\bm{\mathfrak{R}}}\right)}_{i,j}^{n}\right]\right)^{2},
\end{align}

where $N_t$ is the total number of points in time, and all derivatives in equations \eqref{eq:11}, \eqref{eq:22}, and \eqref{eq:33} are evaluated via 1st-order finite-differencing\footnote{The Godunov loss we use is based on the HLLC solver which is 1st-order accurate. For this reason, spatial derivatives used in the PDE-based losses must also be 1st-order accurate to allow for fair comparison.} (n.b. the L2 norms in equations \eqref{eq:11} and \eqref{eq:22} are taken over the space and time dimensions only). $\mathcal{L}_{Visc}$ is an alternative way of obtaining entropy solutions by introducing artificial viscosity. The second term in $\mathcal{L}_{TV+Ent}$ penalizes violations arising from oscillatory behaviour as measured by change in total variation \cite{goodman1985accuracy, krivodonova2021tvd}:

\begin{equation}
\Delta TV = \{TV^{n+1} - TV^{n}\}, \,\,\,\,\,\, TV^{n} := \sum_{i,j}\Delta y \lvert \mathbf{Q}_{i+1,j}^{n} - \mathbf{Q}_{i,j}^{n}\rvert  + \Delta x \lvert \mathbf{Q}_{i,j+1}^{n} - \mathbf{Q}_{i,j}^{n}\rvert.
\end{equation}

The third term penalizes entropy condition violations where $\mathbf{\Tilde{F}} = u\Phi$, $\mathbf{\Tilde{G}} = v\Phi$, and $\Phi := -\rho \mathrm{ln}\left(\frac{p}{\rho^{\gamma}}\right)$.

$\mathcal{L}_{LF}$ is a FVM-based loss that uses Lax-Friedrichs flux estimation for $\mathbf{F}$ and $\mathbf{G}$ in equation \eqref{eq:44}. It is a non-Godunov type loss as the flux estimation does not incorporate solution of the Riemann problem (hence the notation $\cancel{\bm{\mathfrak{R}}}$ in equation \eqref{eq:44}). The Lax-Friedrichs flux is defined in 2D as:

\begin{equation}
    \mathbf{F}_{i + \frac{1}{2}, j}^{n} = 
    \frac{1}{2}\left(\mathbf{F}\left(\mathbf{Q}\right)_{i+1, j}^{n} + \mathbf{F}\left(\mathbf{Q}\right)_{i,j}^{n}\right) - \frac{\Delta x}{2 \Delta t}\left( \mathbf{Q}_{i + 1, j}^{n} -  \mathbf{Q}_{i, j}^{n}\right)
\end{equation}

\begin{equation}
    \mathbf{G}_{i, j+ \frac{1}{2}}^{n} = 
    \frac{1}{2}\left(\mathbf{G}\left(\mathbf{Q}\right)_{i, j + 1}^{n} + \mathbf{G}\left(\mathbf{Q}\right)_{i,j}^{n}\right) - \frac{\Delta y}{2 \Delta t}\left( \mathbf{Q}_{i, j + 1}^{n} -  \mathbf{Q}_{i, j}^{n}\right)
\end{equation}

We perform grid-search and find $\alpha = 0.0075$, $\beta_{1} = 10$, and $\beta_{2} = 1$ give the best results for the regularized PDE losses. We arbitrarily set all four elements of $\bm{\omega}$ to 0.25. We analyse performance of the four losses $\mathcal{L}_G$, $\mathcal{L}_{TV+Ent}$, $\mathcal{L}_{Visc}$, $\mathcal{L}_{LF}$ by comparing their models' predictions with respect to the solution of a reference 5th-order WENO numerical scheme \cite{liu1994weighted}. The WENO ground truths are generated using the JAX-Fluids simulation framework \cite{bezgin2023jax}. The comparison is computed as percentage L2-norm difference between prediction and reference. 

\begin{figure*}[h]
\begin{center}
\centerline{\includegraphics[width=1.00\textwidth]{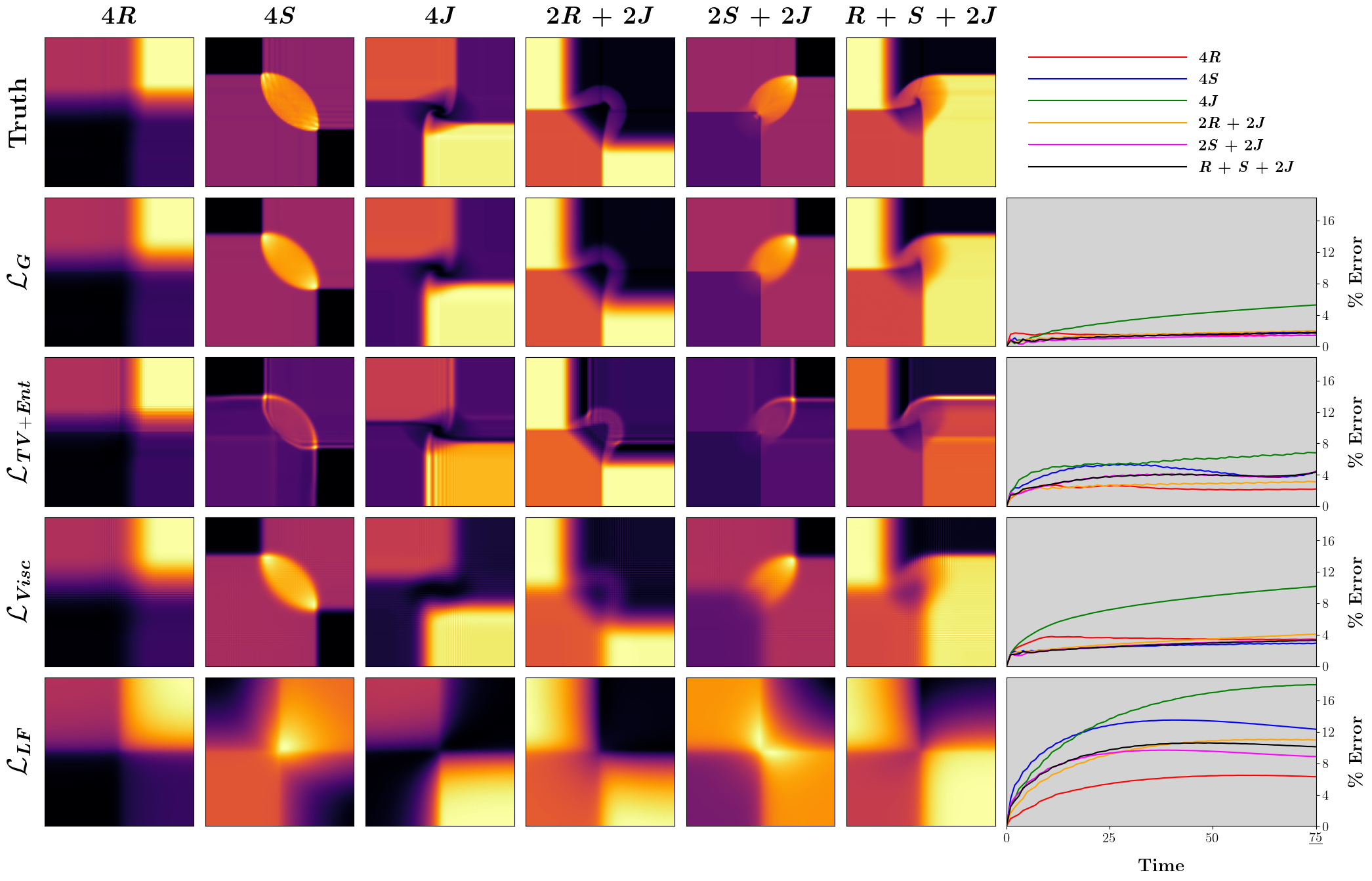}}
\caption{Predicted density fields of Conv-LSTMs trained with losses $\mathcal{L}_G$, $\mathcal{L}_{TV+Ent}$, $\mathcal{L}_{Visc}$ and $\mathcal{L}_{LF}$. Solutions shown are at end of training ($n=75$), for the six configurations of interest. Ground truth solutions are obtained from a 5th-order WENO scheme. The right-most column shows error evolution (with 95\% confidence intervals) of predictions of Conv-LSTMs trained with the different losses. Error values at $n=75$ are outlined in Table \ref{tab:res}.}
\label{fig:contour75}
\end{center}
\vskip -0.3in
\end{figure*}

\begin{figure*}[h]
\begin{center}
\centerline{\includegraphics[width=1.00\textwidth]{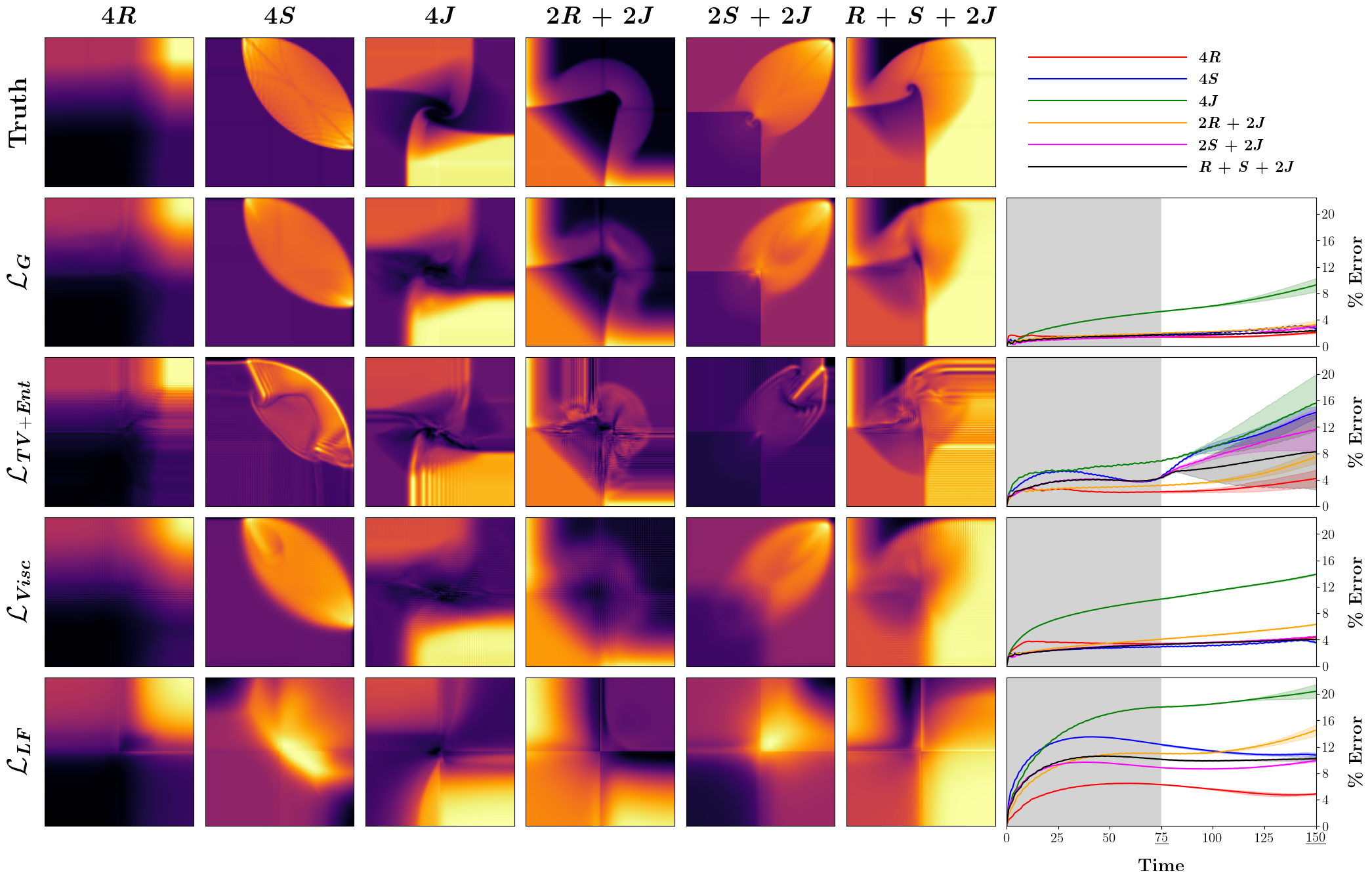}}
\caption{Predicted density fields of Conv-LSTMs trained with losses $\mathcal{L}_G$, $\mathcal{L}_{TV+Ent}$, $\mathcal{L}_{Visc}$ and $\mathcal{L}_{LF}$. Solutions shown are at end of inference ($n=150$), for the six configurations of interest. Ground truth solutions are obtained from a 5th-order WENO scheme. The right-most column shows error evolution (with 95\% confidence intervals) of predictions of Conv-LSTMs trained with the different losses. Shaded regions indicate training range ($n=0-75$). Error values at $n=75$ and $n=150$ are outlined in Table \ref{tab:res}.}
\label{fig:contour}
\end{center}
\vskip -0.3in
\end{figure*}

We observe from Figure \ref{fig:contour75} and Figure \ref{fig:contour} how the Conv-LSTM model trained using loss $\mathcal{L}_G$ gives a visually faithful prediction of configuration $4S$ (i.e., four shock waves) at $n=75$ and $n=150$, with the inference window being double that of the training window. For this configuration Table \ref{tab:res} indicates a mean error of $1.81\%$ at end of training $(n=75)$ and $2.64\%$ at end of inference $(n=150)$, which are significantly lower than the other losses. The prediction using $\mathcal{L}_{Visc}$ also faithfully captures the structure of the solution but in a more diffusive manner. $\mathcal{L}_{TV+Ent}$ on the other hand gives a distorted solution, as does $\mathcal{L}_{LF}$.

Similar trends are observed from Figure \ref{fig:contour} and Table \ref{tab:res} for the other configurations, with the Godunov loss showing superior performance. The disparity in performance between $\mathcal{L}_G$ and the other losses is due to $\mathcal{L}_G$ being more expressive of the problem that is to be solved in the sense that what is expected of the loss (monotonicity, entropy satisfaction) is inherently built into the loss without additional regularizing terms. The Godunov loss is derived from Godunov schemes, which have theoretical guarantees on entropy condition satisfaction and monotonicity. It is therefore expected to outperform the other loss variants. With specific regard to the other losses, a possible reason of the poor performance of $\mathcal{L}_{TV+Ent}$ is its clear tendency to overfit the training range. This overfitting causes distortive artefacts to appear during inference. $\mathcal{L}_{Visc}$ on the other hand is an already diffusive model at the end of training that further propagates diffusive behaviour in time causing loss of information - this is especially the case for $2R + 2J$. Of course a way to decrease this diffusion is by decreasing parameter $\alpha$, but we find that doing this introduces training instability. As for $\mathcal{L}_{LF}$, while its trained model is able to identify the general high-level solution structure during inference, the resemblance of these structures to the reference is remote at best. We show with this example how incorporating a non-Godunov FVM scheme into ML can be insufficient for modelling hyperbolic PDEs. While Godunov losses like $\mathcal{L}_G$ utilize the structure of the Riemann problem so that each intercell flux is estimated from local information, a loss such as $\mathcal{L}_{LF}$ does not. Loss $\mathcal{L}_{LF}$ is based on Lax-Friedrichs flux estimation, which relies on simple averaging and a numerical diffusion term to smear out discontinuities. Such a technique evidently does not lend itself effectively to ML modelling of shocks.

\begin{table*}[tb!]
\caption{Error of Conv-LSTMs trained with $\mathcal{L}_G$, $\mathcal{L}_{TV+Ent}$. $\mathcal{L}_{Visc}$ and $\mathcal{L}_{LF}$ at end of training ($n=75$) and at end of inference ($n=150$), for the six configurations of interest. Errors (reported with 95\% confidence intervals) are computed as percentage L2-norm difference between ML predictions and WENO references.}
\begin{center}
\begin{small}
\begin{sc}
\resizebox{1.0\textwidth}{!}{\begin{tabular}{ccccccccc}
\toprule
\multirow{2}{*}{\textbf{Con}} \hspace{0.0mm} & \multicolumn{4}{c}{\textbf{\% Error $(n = \underline{75})$}}\hspace{0.0mm} &  \multicolumn{4}{c}{\textbf{\% Error $(n = \underline{150})$}}\vspace{1mm} \\ \cmidrule(lr){2-5} \cmidrule(l){6-9}
& $\mathcal{L}_{G}$ & $\mathcal{L}_{TV+Ent}$ & $\mathcal{L}_{Visc}$ & $\mathcal{L}_{LF}$ \hspace{0.0mm} & $\mathcal{L}_{G}$ & $\mathcal{L}_{TV+Ent}$ & $\mathcal{L}_{Visc}$ & $\mathcal{L}_{LF}$\\
\midrule
$4R$ \hspace{0.0mm} & $\mathbf{1.39 \pm 0.02}$ & $2.19 \pm 0.17$ & $3.42\pm 0.03$ & $6.29\pm 0.01$ \hspace{0.0mm} & $\mathbf{2.09\pm 0.16}$ & $4.20\pm 1.31$ & $4.37\pm 0.13$ & $4.88\pm 0.14$\\
$4S$ \hspace{0.0mm} & $\mathbf{1.81\pm 0.03}$ & $4.46\pm 0.00$ & $2.93\pm 0.01$ & $12.34\pm 0.13$ \hspace{0.0mm} & $\mathbf{2.64\pm 0.14}$ & $14.18\pm 0.94$ & $3.56\pm 0.10$ & $10.75\pm 0.33$\\
$4J$ \hspace{0.0mm} & $\mathbf{5.26\pm 0.02}$ & $6.80\pm 0.03$ & $10.16\pm 0.02$ & $18.00\pm 0.04$ \hspace{0.0mm} & $\mathbf{9.28\pm 1.00}$ & $15.59\pm 4.28$ & $13.91\pm 0.07$ & $20.40\pm 1.04$\\
$2R + 2J$ \hspace{0.0mm} & $\mathbf{1.96\pm 0.02}$ & $3.11\pm 0.06$ & $4.07\pm 0.01$ & $10.95\pm 0.02$ \hspace{0.0mm} & $\mathbf{3.27\pm 0.64}$ & $7.44\pm 0.99$ & $6.32\pm 0.05$ & $14.50\pm 0.84$\\
$2S + 2J$ \hspace{0.0mm} & $\mathbf{1.42\pm 0.00}$ & $4.41\pm 0.21$ & $3.33\pm 0.02$ & $8.85\pm 0.01$ \hspace{0.0mm} & $\mathbf{3.05\pm 0.21}$ & $11.59\pm 3.17$ & $4.49\pm 0.08$ & $9.91\pm 0.16$\\
$R + S + 2J$ \hspace{0.0mm} & $\mathbf{1.67\pm 0.03}$ & $4.33\pm 0.10$ & $3.28\pm 0.00$ & $10.11\pm 0.08$ \hspace{0.0mm} & $\mathbf{2.39\pm 0.16}$ & $8.22\pm 5.76$ & $4.08\pm 0.14$ & $10.21\pm 0.34$\\
\bottomrule
\end{tabular}}\label{tab:res}
\end{sc}
\end{small}
\end{center}
\vskip -0.1in
\end{table*}

Figure \ref{fig:contour} and Table \ref{tab:res} indicate that $\mathcal{L}_G$ struggles most with the case of four contact waves ($4J$). This could be due to the solution exhibiting a larger range of length scales compared to other configurations, because of the vorticity that evolves from the centre of the domain.  As a result the model struggles to capture the finer features of the vorticity towards the centre. Configuration $R+S+2J$ also appears to have this problem but to a lesser extent. A possible solution is to incorporate multiscale sub-architectures into the model.

\subsection{Inverse Problem: Super-Resolution Reconstruction}

We examine next the performance of the Godunov loss in the super-resolution reconstruction of the 2D Riemann problem. Here we focus on a single configuration and try to reconstruct it from three levels of low resolution. The configuration is another $4S$ variant, $S_{21}^{-}S_{32}^{-}S_{34}^{-}S_{41}^{-}$, in the domain $\left[0.3,0.7\right] \times \left[0.3,0.7\right]$. See \ref{app:convalues} for initialization values of this configuration.

To generate the low resolution input, we numerically generate a $256^2$ WENO solution and average-pool two subsequent snapshots by factors $\times4$, $\times8$, and $\times16$. We need two snapshots, spaced $\Delta t = 0.00005s$, as a minimum to compute the time-derivative in the physical losses. We then use the VDSR-based model (Figure \ref{fig:superresdia}) to perform the reconstruction\footnote{Training is performed on a single NVIDIA A100 GPU, requiring up to $\sim$ 4 Gb of RAM, as well as $\sim$ 20 minutes of compute time.}. The input passes through two convolution-ReLU layers followed by a convolution-interpolation layer, where interpolation is done bilinearly. This is repeated until the desired resolution of $256^2$ is reached. This upscaled representation is then passed into the VDSR module where it undergoes eighteen convolution-ReLU layers followed by a residual connection with the input to the VDSR module. The number of hidden channels is 256 for the upsampling module and 64 for the VDSR module, and once again a kernel size of 3 is used\footnote{Tuning of hyperparameters is performed through a random grid-search of different combinations of hyperparameters.}. The model is Xavier-initialized and trained with an Adam optimizer at a learning rate of $5\times10^{-5}$.

\begin{figure*}[h]
\begin{center}
\centerline{\includegraphics[width=1.00\textwidth]{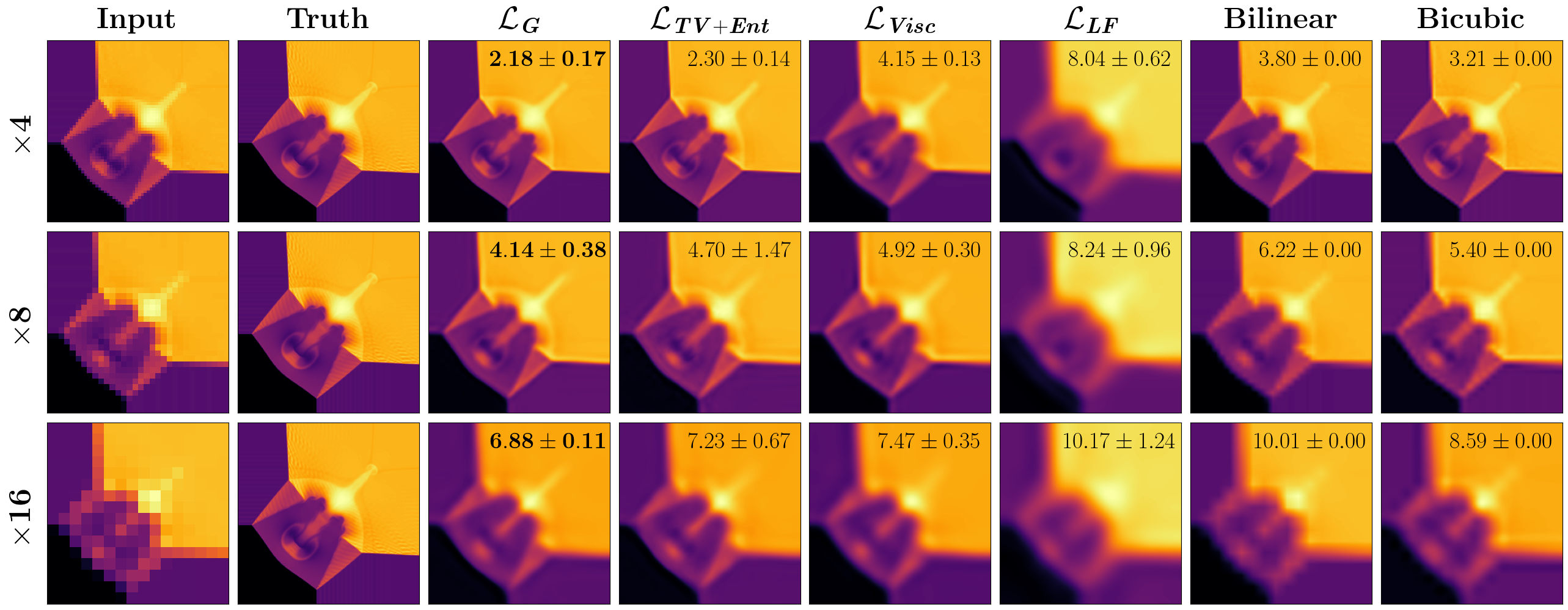}}
\caption{Super-resolution reconstruction of a 2D Riemann problem using VDSR-based models trained with losses $\mathcal{L}_G$, $\mathcal{L}_{TV+Ent}$, $\mathcal{L}_{Visc}$ and $\mathcal{L}_{LF}$. Bilinear and bicubic interpolation outputs are also included for additional perspective. Each row shows super-resolution from a different input resolution. Errors (reported with 95\% confidence intervals) on the top right of each plot are with respect to the WENO truth.}
\label{fig:super}
\end{center}
\vskip -0.3in
\end{figure*}

We again compare performance of $\mathcal{L}_G$ (equation \eqref{eq:godloss}) against $\mathcal{L}_{TV+Ent}$, $\mathcal{L}_{Visc}$ and $\mathcal{L}_{LF}$ (equations \eqref{eq:22} - \eqref{eq:44}) in terms of error difference with respect to the high resolution WENO truth. We find it is necessary for all the losses to have a regularization on solution boundedness. Given input $\mathbf{X}_{LR}$ and prediction $\mathbf{X}_{HR}$ this regularization takes the form:

\begin{equation}
    R := \lambda\norm{\mathcal{AP}\left(\mathbf{X}_{HR}\right) - \mathbf{X}_{LR}},
\end{equation}

where $\mathcal{AP}$ denotes average-pooling and $\lambda$ is the regularization weighting, which we set to 25 for all cases. Figure \ref{fig:super} examines performance of the different losses and also includes outputs from bilinear and bicubic interpolation. For the same reasoning as in the time-stepping problem, $\mathcal{L}_G$ provides the most accurate reconstruction both visually and in terms of $\%$ error at all downsampling levels. $\mathcal{L}_{TV+Ent}$ performs similarly well to $\mathcal{L}_G$ albeit with a slight decrease in accuracy. The outputs using $\mathcal{L}_{Visc}$ are visually adequate at all downsampling levels but do have noticeable blur. $\mathcal{L}_{LF}$ exhibits the worst performance with significant blurring. Bilinear and bicubic interpolation give visually faithful outputs at $\times 4$ but produce checkerboard artifacts along discontinuities at $\times 8$ and $\times 16$, with errors consistently higher than that of $\mathcal{L}_G$ and $\mathcal{L}_{TV+Ent}$.

\begin{figure*}[h]
\begin{center}
\centerline{\includegraphics[width=1.00\textwidth]{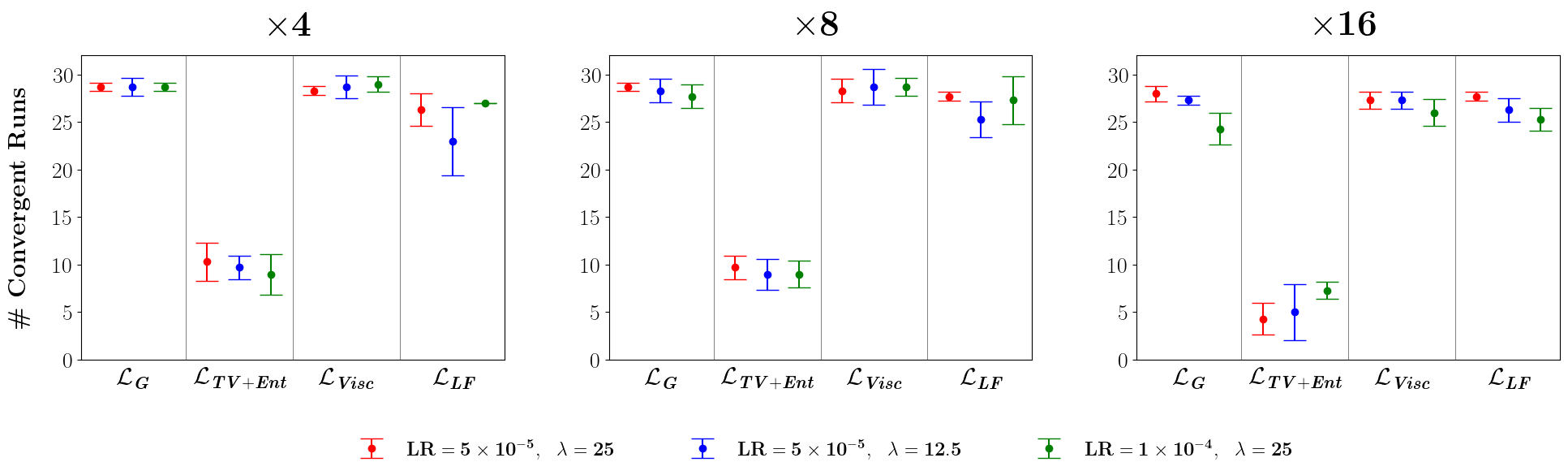}}
\caption{Number of converging runs (per 30 runs) for models trained with losses $\mathcal{L}_G$, $\mathcal{L}_{TV+Ent}$, $\mathcal{L}_{Visc}$ and $\mathcal{L}_{LF}$. Subplots show convergence behaviour at $\times 4$, $\times 8$, $\times 16$ lower resolution levels, respectively. Each color represents a different choice of learning rate and regularization parameter $\lambda$. Error bars show $\pm 1$ standard deviation.}
\label{fig:conv}
\end{center}
\vskip -0.3in
\end{figure*}

It is interesting to also compare the convergence behaviour of all four losses in performing the reconstruction. To do this, we train the VDSR-based model from scratch until epoch 100, and repeat this procedure thirty times. We do this for each loss, for each sub-resolution level, and for three different random seeds, to determine the number of runs (per thirty runs) that are deemed 'towards convergence' within $5000$ epochs. To determine whether a run is on the path to convergence within $5000$ epochs, we take the reconstructed output at epoch 100 and calculate its error with respect to the truth. If this error is within $30\%$ of the smallest error found at epoch 100 (over the 30 runs), then we assume the training is stable and converging. We find that this criterion is consistent with (and more systematic than) deciding whether training will converge based on visual output inspection at epoch 100. We plot the number of converging runs as in Figure \ref{fig:conv}, for different values of learning rate and regularization parameter $\lambda$. From the figure it is clear that $\mathcal{L}_{TV + Ent}$ struggles with convergence in all parts of the parameter space, whilst $\mathcal{L}_{G}$, $\mathcal{L}_{Visc}$, and $\mathcal{L}_{LF}$ are mostly converging.


\section{Conclusion}\label{sec:conc}

In this paper we examine a Godunov loss function for modelling hyperbolic PDEs. We use this loss with a Conv-LSTM architecture to predict the evolution of the 2D Riemann problem under the influence of the 2D Euler equations. We also use the loss with a VDSR-based architecture to reconstruct the 2D Riemann problem from very low resolutions. Results show high accuracy with respect to a numerical reference and a superior, well-rounded performance compared to PDE-based losses with regularizations and compared to a non-Godunov FVM-based loss. The superior performance is due to the Godunov loss being inherently expressive of entropy and monotonicity constraints. The Godunov loss requires minimal hyperparameter tuning simplifying the search space for an optimized ML model. An obvious next step for this work is to extend to the 3D Riemann problem or to approach other classical problems such as high-speed compressible flow past a cylinder. There is also scope to improve training stability of models that use Godunov losses as well as explore the use of adaptive activation functions.

Finally, we emphasize that the novelty of the paper is in the loss function and not in the architecture itself. We provide some justifications as to why the architectures we use are desirable, but ultimately the choice of architectures is arbitrary in that there are numerous other architectures capable of performing the same forward and inverse tasks. Therefore any choice of architecture can be coupled with the Godunov loss that we devise. If, for example, the simulation data is unstructured, then we can combine our loss with an architecture that inherently handles unstructured data, such as graph-based networks.

\section*{Code and Data}

Code and data can be accessed via the
\href{https://github.com/RamiCassia/Godunov-Loss-Functions-for-Modelling-of-Hyperbolic-Conservation-Laws}{{\textcolor{blue}{following link}}}.

Additional information about data generation using JAX-Fluids can be accessed via the \href{https://github.com/tumaer/JAXFLUIDS}{{\textcolor{blue}{following link}}} \cite{bezgin2023jax}.
\newpage
\bibliography{bibli}


\newpage
\appendix


\section{2D Riemann Configurations}\label{app:convalues}
\subsection{Thermodynamic Relations}
A unit square domain, typically $\left[0,1\right] \times \left[0,1\right]$, is divided into four quadrants where each quadrant is uniformly initialized with its own set of $\mathbf{X} = \left(\rho, u, v, p\right)^{T}$. The quadrants are numbered 1 to 4, anticlockwise, starting from the top-right quadrant and finishing in the bottom-right quadrant. The initialization of the quadrants determines the types of the four waves separating them. We denote the waves by their positions based on quadrant numbering, i.e., $lr = \{21, 32, 34, 41\}$. There are three wave types - rarefaction $R$, shock $S$, or contact $J$, characterized by their thermodynamic relations across the wave \cite{schulz1993classification}. In the following, let $w$ and $w^{'}$ denote the normal and tangential velocities (w.r.t waves).

The relations for rarefaction waves ${R}_{lr}^{\pm}$ are as follows:

\begin{align}
& w_{l} - w_{r} = \pm \frac{2\sqrt{\gamma}}{\gamma - 1}\left( \sqrt{\frac{p_l}{\rho_l}} - \sqrt{\frac{p_r}{\rho_r}} \right),\\[10pt]
&\frac{\rho_l}{\rho_r} = \left(\frac{p_l}{p_r}\right)^{\frac{1}{\gamma}},\\[10pt]
& w^{'}_{l} = w^{'}_{r}, \,\,\,\,\,\,\,\, w_l < w_r, \,\,\,\,\,\,\,\,sgn\left( p_{l} - p_{r}\right) = \mp 1.
\end{align}

The relations for shock waves ${S}_{lr}^{\pm}$ are as follows:

\begin{align}
&w_{l} - w_{r} = \sqrt{\frac{\left(p_{l} - p_{r}\right)\left(\rho_{l} - \rho_{r}\right)}{\rho_l \rho_r}}, \\[10pt]
& \frac{\rho_{l}}{\rho_{r}} = \frac{\gamma\left(p_{l} + p_{r}\right) + \left(p_{l} - p_{r}\right)}{\gamma\left(p_{l} + p_{r}\right) - \left(p_{l} - p_{r}\right)},\\[10pt]
& w^{'}_{l} = w^{'}_{r},\,\,\,\,\,\,\,\, w_l > w_r,\,\,\,\,\,\,\,\, sgn\left( p_{l} - p_{r}\right) = \pm 1.
\end{align}

The relations for contact discontinuities ${J}_{lr}^{\pm}$ are as follows:

\begin{align}
&sgn\left( w^{'}_{l} - w^{'}_{r}\right) = \mp 1, \,\,\,\,\,\,\,\,
    w_{l} = w_{r}, \,\,\,\,\,\,\,\, p_{l} = p_{r}.
\end{align}


Note that energy $E$ is defined as:

\begin{equation}
E := \frac{1}{2}\rho\left(u^2 + v^2\right) + \frac{p}{\left(\gamma - 1\right)}.
\end{equation}

\newpage
\subsection{Initialization Values}
The following are the initialization values of the 2D Riemann configurations explored in this work. For initialization values of other 2D Riemann configurations, see \citet{kurganov2002solution}.

\begin{alignat*}{2}
&R_{21}^{+}R_{32}^{+}R_{34}^{+}R_{41}^{+} &&: =\begin{cases}
    \mathbf{X}_1 = \left(1.0000, \:\:\:\: 0.0000, \:\:\:\: 0.0000, \:\:\:\: 1.0000 \right), & \text{$x>0.5$, \,\,\,\, $y>0.5$}\\
    \mathbf{X}_2 = \left(0.5197, \, -0.7259, \:\:\:\: 0.0000, \:\:\:\: 0.4000 \right), & \text{$x<0.5$, \,\,\,\, $y>0.5$}\\
    \mathbf{X}_3 = \left(0.1072, \,  -0.7259, \, -1.4045, \:\:\:\: 0.0439 \right), & \text{$x<0.5$, \,\,\,\, $y<0.5$}\\
    \mathbf{X}_4 = \left(0.2579, \:\:\:\:  0.0000, \, -1.4045, \:\:\:\: 0.1500 \right), & \text{$x>0.5$, \,\,\,\, $y<0.5$}
  \end{cases} \\ \\
&S_{21}^{-}S_{32}^{-}S_{34}^{-}S_{41}^{-}&&: =\begin{cases}
    \mathbf{X}_1 = \left(1.5000, \:\:\:\: 0.0000, \:\:\:\: 0.0000, \:\:\:\: 1.5000 \right), & \text{$x>0.5$, \,\,\,\, $y>0.5$}\\
    \mathbf{X}_2 = \left(0.5323, \:\:\:\: 1.2060, \:\:\:\: 0.0000, \:\:\:\: 0.3000 \right), & \text{$x<0.5$, \,\,\,\, $y>0.5$}\\
    \mathbf{X}_3 = \left(0.1380, \:\:\:\: 1.2060,\:\:\:\: 1.2060, \:\:\:\: 0.0290 \right), & \text{$x<0.5$, \,\,\,\, $y<0.5$}\\
    \mathbf{X}_4 = \left(0.5323, \:\:\:\: 0.0000, \:\:\:\: 1.2060,\:\:\:\: 0.3000\right), & \text{$x>0.5$, \,\,\,\, $y<0.5$}
  \end{cases} \\ \\
&S_{21}^{-}S_{32}^{+}S_{34}^{+}S_{41}^{-}&&: =\begin{cases}
    \mathbf{X}_1 = \left(1.1000, \:\:\:\: 0.0000, \:\:\:\: 0.0000, \:\:\:\: 1.1000 \right), & \text{$x>0.5$, \,\,\,\, $y>0.5$}\\
    \mathbf{X}_2 = \left(0.5065, \:\:\:\: 0.8939, \:\:\:\: 0.0000, \:\:\:\: 0.3500 \right), & \text{$x<0.5$, \,\,\,\, $y>0.5$}\\
    \mathbf{X}_3 = \left(1.1000, \:\:\:\: 0.8939,\:\:\:\: 0.8939, \:\:\:\: 1.1000 \right), & \text{$x<0.5$, \,\,\,\, $y<0.5$}\\
    \mathbf{X}_4 = \left(0.5065, \:\:\:\: 0.0000, \:\:\:\: 0.8939,\:\:\:\: 0.3500\right), & \text{$x>0.5$, \,\,\,\, $y<0.5$}
  \end{cases} \\ \\
&J_{21}^{-}J_{32}^{+}J_{34}^{-}J_{41}^{+}&&: =\begin{cases}
    \mathbf{X}_1 = \left( 1.0000,\:\:\:\: 0.7500, \, -0.5000,  \:\:\:\: 1.0000  \right), & \text{$x>0.5$, \,\,\,\, $y>0.5$}\\
    \mathbf{X}_2 = \left(2.0000, \:\:\:\:  0.7500,\:\:\:\: 0.5000, \:\:\:\: 1.0000 \right), & \text{$x<0.5$, \,\,\,\, $y>0.5$}\\
    \mathbf{X}_3 = \left(1.0000, \, -0.7500,\:\:\:\: 0.5000, \:\:\:\: 1.0000 \right), & \text{$x<0.5$, \,\,\,\, $y<0.5$}\\
    \mathbf{X}_4 = \left(3.0000, \, -0.7500, \,-0.5000, \:\:\:\: 1.0000\right), & \text{$x>0.5$, \,\,\,\, $y<0.5$}
  \end{cases} \\ \\ 
&R_{21}^{-}J_{32}^{-}J_{34}^{-}R_{41}^{-}&&: =\begin{cases}
    \mathbf{X}_1 = \left(0.5197, \:\:\:\:0.1000, \:\:\:\:0.1000,  \:\:\:\:0.4000  \right), & \text{$x>0.5$, \,\,\,\, $y>0.5$}\\
    \mathbf{X}_2 = \left(1.0000, \, -0.6259, \:\:\:\:0.1000,  \:\:\:\:1.0000  \right), & \text{$x<0.5$, \,\,\,\, $y>0.5$}\\
    \mathbf{X}_3 = \left(0.8000, \:\:\:\: 0.1000,\:\:\:\: 0.1000, \:\:\:\: 1.0000 \right), & \text{$x<0.5$, \,\,\,\, $y<0.5$}\\
    \mathbf{X}_4 = \left(1.0000, \:\:\:\: 0.1000, \, -0.6259, \:\:\:\: 1.0000 \right), & \text{$x>0.5$, \,\,\,\, $y<0.5$}
  \end{cases}\\ \\
&S_{21}^{+}J_{32}^{+}J_{34}^{+}S_{41}^{+}&&: =\begin{cases}
    \mathbf{X}_1 = \left(0.5313, \:\:\:\:0.0000, \:\:\:\:0.0000,  \:\:\:\:0.4000  \right), & \text{$x>0.5$, \,\,\,\, $y>0.5$}\\
    \mathbf{X}_2 = \left(1.0000, \:\:\:\: 0.7276, \:\:\:\:0.0000,  \:\:\:\:1.0000  \right), & \text{$x<0.5$, \,\,\,\, $y>0.5$}\\
    \mathbf{X}_3 = \left(0.8000, \:\:\:\: 0.0000, \:\:\:\:0.0000,  \:\:\:\:1.0000  \right), & \text{$x<0.5$, \,\,\,\, $y<0.5$}\\
    \mathbf{X}_4 = \left(1.0000, \:\:\:\: 0.0000, \:\:\:\:0.7276, \:\:\:\: 1.0000\right), & \text{$x>0.5$, \,\,\,\, $y<0.5$}
  \end{cases} \\ \\
&R_{21}^{-}J_{32}^{-}J_{34}^{+}S_{41}^{+} &&: =\begin{cases}
    \mathbf{X}_1 = \left( 0.5313,\:\:\:\:0.1000, \:\:\:\:0.1000,  \:\:\:\:0.4000  \right), & \text{$x>0.5$, \,\,\,\, $y>0.5$}\\
    \mathbf{X}_2 = \left( 1.0222,\,-0.6179,  \:\:\:\:0.1000,  \:\:\:\:1.0000  \right), & \text{$x<0.5$, \,\,\,\, $y>0.5$}\\
    \mathbf{X}_3 = \left(0.8000, \:\:\:\:0.1000, \:\:\:\: 0.1000, \:\:\:\:1.0000  \right), & \text{$x<0.5$, \,\,\,\, $y<0.5$}\\
    \mathbf{X}_4 = \left(1.0000, \:\:\:\:0.1000,  \:\:\:\:0.8276, \:\:\:\: 1.0000\right), & \text{$x>0.5$, \,\,\,\, $y<0.5$}
  \end{cases}
\end{alignat*}   







\end{document}